\documentclass[aps,prd,a4paper,preprintnumbers,superscriptaddress,twocolumn,floatfix,showpacs,nofootinbib]{revtex4}
\usepackage{amssymb}
\usepackage{amsmath}

\usepackage{float,graphicx}

\begin{document}

\title{Testing Alternative Theories of Dark Matter with the CMB}
\author{Baojiu~Li}
\email[Email address: ]{b.li@damtp.cam.ac.uk}
\affiliation{DAMTP, Centre for Mathematical Sciences, University of Cambridge, Cambridge
CB3 0WA, United Kingdom}
\author{John~D.~Barrow}
\email[Email address: ]{j.d.barrow@damtp.cam.ac.uk}
\affiliation{DAMTP, Centre for Mathematical Sciences, University of Cambridge, Cambridge
CB3 0WA, United Kingdom}
\author{David~F.~Mota}
\email[Email address: ]{d.mota@thphys.uni-heidelberg.de}
\affiliation{Institute of Theoretical Physics, University of Heidelberg, 69120
Heidelberg, Germany}
\author{HongSheng~Zhao}
\email[Email address: ]{hz4@st-andrews.ac.uk}
\affiliation{Scottish University Physics Alliance, University of St.~Andrews, KY16 9SS,
United Kingdom}
\date{\today}

\begin{abstract}
We propose a method to study and constrain modified gravity theories for
dark matter using CMB temperature anisotropies and polarization. We assume
that the theories considered here have already passed the matter
power-spectrum test of large-scale structure. With this requirement met, we
show that a modified gravity theory can be specified by parametrizing the
time evolution of its dark-matter density contrast, which is completely
controlled by the dark matter stress history. We calculate how the stress
history with a given parametrization affects the CMB observables, and a
qualitative discussion of the physical effects involved is supplemented with
numerical examples. It is found that, in general, alternative gravity
theories can be efficiently constrained by the CMB temperature and
polarization spectra. There exist, however, special cases where modified
gravity cannot be distinguished from the CDM model even by using \emph{both}
CMB \emph{and} matter power spectrum observations, nor can they be
efficiently restricted by other observables in perturbed cosmologies. Our
results show how the stress properties of dark matter, which determine the
evolutions of both density perturbations and the gravitational potential,
can be effectively investigated using just the general conservation
equations and without assuming any specific theoretical gravitational theory
within a wide class.
\end{abstract}

\maketitle

\section{Introduction}

\label{sect:Introduction}

Various cosmological observations indicate that almost $95\%$ of the energy
content of our Universe is in some dark form that can only be detected
through its gravitational effects. The standard explanation for this "dark
sector" involves some kind of cold dark matter (CDM), such as heavy weakly
interacting particles or primordial black holes, and dark energy, which
could be a manifestation of a cosmological constant ($\Lambda $) or some
exotic matter field. This standard hybrid picture ($\Lambda \mathrm{CDM}$)
has so far passed several different cosmological tests, and is dubbed the
"concordance model". Despite of its successes, however, it is not without
problems in accounting for structure on galactic scales -- for example, the
greatest challenge is to form bulge-free bright spiral galaxies and dwarf
spiral galaxies. The galaxies in CDM have either a dominating stellar bulge
or a dominating CDM cusp at the center. In essence LCDM over-predicts the
dark-matter effects required by the empirical Tully-Fisher relation of
spiral galaxies -- and is in need of an experimental identification of the
CDM particles with an understanding why the dark energy possesses its
particular energy density.

Since dark energy affects cosmological models mainly via gravitational
effects, it is possible to imagine that the effects of an explicit material
source for the dark energy can be mimicked by a change in the behaviour of
the gravitational field at late times. This is an unusual requirement since
we have expected deviations from general relativity to arise in the high
spacetime curvature limit at early times rather than in the low spacetime
curvature limit at late times. However, we should note that the addition of
an explicit cosmological constant to general relativity, as in the
concordance model of $\Lambda \mathrm{CDM}$, is already a particular example
of such a low spacetime curvature correction. There have been many
investigations of gravitational alternatives to dark energy \cite%
{Carroll2005, Easson2004, Vollick2003, brook, Allemandi2004a,
Allemandi2004b, Nojiri2005, amar,Nojiri2006}, and it appears that most of
these attempts create different problems in either local gravitational
systems or large-scale structure (LSS) formation, or even both \cite%
{Chiba2003, Erickcek2006, Navarro2007, Faulkner2007, Barausse2008, Hu2007a,
Olmo2005, Koivisto2006, Li2006, Li2007a, Li2007b, Li2007c, Li2007d,
Amendola2007, Li2008a, Li2008b} (see \cite{hu, doug1, doug2, cham, apple,
amen} on possible ways to overcome these problems). In response to these
investigations, frameworks to test modified gravitational dark energy models
have also been developed \cite{Jain2007, hans, Hu2007b, Bertschinger2008,
Hu2008a}.

The situation is nonetheless different in the dark matter arena, where the
leading modified gravity model, Milgrom's modified Newtonian dynamics (MOND)
\cite{mond}, appeared more than two decades ago, but lacked a convincing
general-relativistic formulation with a set of relativistic field equations
that are applicable to cosmology. This hurdle has recently been overcome by
Bekenstein's tensor-vector-scalar (TeVeS) model \cite{TeVeS} which reduces
to MOND in the relevant limit. Actually, what the TeVeS model provides is
more than a relativistic framework to investigate cosmology -- it has been
found that in TeVeS the formation of LSS, which was thought of as a problem
for modified gravity theories before, could also be made consistent with
observations \cite{TeVeSperturbation, TeVeSVector}. Meanwhile, TeVeS has
also been shown to work well on smaller scales (\emph{e.g.}, mimicking cold
dark matter in strong gravitational lensing systems \cite{Zhao2006, Shan2008}%
, producing elliptical galaxies, barred spiral galaxies and even tidal dwarf
galaxies \cite{Tiret2007, Gentile2007}, being consistent with solar system
tests and so on) and inherit the advantages of MOND over the CDM model \cite%
{Gnedin2002} (\emph{e.g.}, in explaining the Tully-Fisher law and galaxy
rotation curves), this discovery attracts much interest on TeVeS and its
generalizations. Also, since TeVeS manages to grow large-scale structure
using the growing mode of its vector field, this stimulates the
investigation of vector-field cosmology \cite{Jacobson} in general \cite%
{TeVeSAether, Ferreira2006, Zlosnik2007a, Li2008c, tomi1, fran, Zhao2007,
Skordis2008, Zhao2008} (in Ref.~\cite{Zhao2008}, for example, a very general
vector-scalar field framework, the so-called dark fluid, is proposed which
can reduce to many existing models in appropriate limits), and it is found
that the LSS in these theories could also be consistent with observations
\cite{Zlosnik2007b}.

Despite of this encouraging success, one should bear in mind that LSS only
provides one test of any structure formation model. In fact, the matter (or
galaxy) power spectrum we observe today only reflects the large-scale
matter-density perturbation \emph{today} ($\delta _{m0}$), rather than its
evolutionary history. Thus, even though the matter power spectrum, $P(k),$
is compatible with observations, the evolution path of $\delta _{m}$ may
well be different from that followed in the $\Lambda \mathrm{CDM}$ paradigm.
This situation is shown particularly clearly in Fig.~3 (lower panel) of \cite%
{TeVeSVector}, where the growth rate of baryon density perturbations is
enhanced only at late times. The different evolution history of
matter-density perturbations may have significant impacts on various other
cosmological observables, such as the cosmic microwave background (CMB)
power spectrum, and influence the growth of nonlinear structure. We would
like to determine what these impacts are.

In this paper we take a first step in that direction. We will concentrate on
the influences of general modified gravity dark-matter models on CMB
observables. Our main assumption is that the modification to general
relativity (GR) can be expressed as an \textit{effective dark matter} (EDM)
term and moved to the right-hand side of the field equations so that the
left-hand side is the same as in GR. This EDM term, like the standard CDM,
governs both the background and perturbation evolutions. This assumption is
justified by observing the fact that in TeVeS, as well as in the general
vector-field (or $f(K)$) theories, the terms involving the vector field are
essentially just the EDM term described here. Furthermore, we make some
simplifying assumptions. First, any explicit dark energy is neglected. The
main effects of dark energy (if not too exotic in origin) are to modify the
background evolution at late times and cause the decay of the large-scale
gravitational potential. Neglecting it does not affect the essential
features of our model but will simplify the numerics greatly. Second, the
background evolution is exactly the same as that in the standard CDM (SCDM)
model, which should also be a good approximation. Third, the matter power
spectrum, or equivalently $\delta _{m0}$, is the same as that of SCDM,
because any deviation from the latter should be stringently constrained by
LSS observations, and because both TeVeS and $f(K)$ models have claimed to
reproduce the observed matter power spectrum. Hence, we are fixing $\delta
_{m0}$ and investigating how different evolutions of $\delta _{m}(a)$ (where
$a$ is the cosmic scale factor) affect the CMB power spectrum. Our
theoretical framework is designed to be more general than is required for
this purpose alone, and could be used to investigate features of other
cosmological models.

This paper is organized as follows: in \S ~\ref{sect:Model} we set out the
theoretical framework for our investigation and introduce more details of
the cosmological model. In \S ~\ref{sect:CMB} we describe briefly how a
general dark-matter component affects the CMB power spectrum and then, in \S %
~\ref{sect:Numerics}, supplement this discussion with a numerical example.
We consider three special cases which span all the possibilities in the
model, and explain them one by one. Finally, in \S ~\ref{sect:Conclusion} we
provide a summary of our results together with some further discussion of
the assumptions on which they are based.

\section{The Theoretical Framework}

\label{sect:Model}

In this analysis we use the perturbed Einstein equations in the
covariant and gauge invariant (CGI) formalism.

\subsection{The Perturbation Equations in CGI Formalism}

\label{sect:appendixa}

The CGI perturbation equations in general \AE -theories are
derived in this section using the method of $3+1$ decomposition
\cite{GR3+1}. First, we briefly review the main ingredients of
$3+1$ decomposition and their application to standard general
relativity \cite{GR3+1} for ease of later reference.

The main idea of $3+1$ decomposition is to make spacetime splits
of physical quantities with respect to the 4-velocity $u^{a}$ of
an observer. The projection tensor $h_{ab}$ is defined as
$h_{ab}=g_{ab}-u_{a}u_{b}$ and can be used to obtain covariant
tensors perpendicular to $u$. For example, the covariant spatial
derivative $\hat{\nabla}$ of a tensor field $T_{d\cdot \cdot \cdot
e}^{b\cdot \cdot \cdot c}$ is defined as
\begin{eqnarray}
\label{eq:AEEOM} \hat{\nabla}^{a}T_{d\cdot \cdot \cdot e}^{b\cdot
\cdot \cdot c}\equiv h_{i}^{a}h_{j}^{b}\cdot \cdot \cdot \
h_{k}^{c}h_{d}^{r}\cdot \cdot \cdot \ h_{e}^{s}\nabla
^{i}T_{r\cdot \cdot \cdot s}^{j\cdot \cdot \cdot k}.
\end{eqnarray}%
The energy-momentum tensor and covariant derivative of the
4-velocity are decomposed respectively as
\begin{eqnarray}
\label{eq:AEEMT} T_{ab} &=&\pi _{ab}+2q_{(a}u_{b)}+\rho u_{a}u_{b}-ph_{ab}, \\
\nabla _{a}u_{b} &=&\sigma _{ab}+\varpi _{ab}+\frac{1}{3}\theta
h_{ab}+u_{a}A_{b}.
\end{eqnarray}%
In the above, $\pi _{ab}$ is the projected symmetric trace-free
(PSTF) anisotropic stress, $q_{a}$ the vector heat flux vector,
$p$ the isotropic pressure, $\sigma _{ab}$ the PSTF shear tensor,
$\varpi _{ab}=\hat{\nabla}_{[a}u_{b]}$, the vorticity, $\theta
=\nabla ^{c}u_{c}\equiv 3\dot{a}/a$ ($a$ is the mean expansion
scale factor) the expansion scalar, and $A_{b}=\dot{u}_{b}$ the
acceleration; the overdot denotes time derivative expressed as
$\dot{\phi}=u^{a}\nabla _{a}\phi $, brackets mean
antisymmetrisation, and parentheses symmetrization. The 4-velocity
normalization is chosen to be $u^{a}u_{a}=1$. The quantities
$\pi_{ab},q_{a},\rho ,p$ are referred to as \emph{dynamical}
quantities and $\sigma _{ab},\varpi _{ab},\theta ,A_{a}$ as
\emph{kinematical} quantities. Note that the dynamical quantities
can be obtained from the energy-momentum tensor $T_{ab}$ through
the relations
\begin{eqnarray}
\rho &=&T_{ab}u^{a}u^{b},  \nonumber \\
p &=&-\frac{1}{3}h^{ab}T_{ab},  \nonumber \\
q_{a} &=&h_{a}^{d}u^{c}T_{cd},  \nonumber \\
\label{eq:DefDynamicalQuantity} \pi _{ab}
&=&h_{a}^{c}h_{b}^{d}T_{cd}+ph_{ab}.
\end{eqnarray}

Decomposing the Riemann tensor and making use the Einstein
equations, we obtain, after linearization, five constraint
equations \cite{GR3+1}:
\begin{eqnarray}
\label{eq:ConstraintVarpi} 0 &=&\hat{\nabla}^{c}(\varepsilon _{\ \ cd}^{ab}u^{d}\varpi _{ab}); \\
\label{eq:Constraintq} \kappa q_{a} &=&
-\frac{2\hat{\nabla}_{a}\theta}{3} +
\hat{\nabla}^{b}\sigma_{ab}+\hat{\nabla}^{b}\varpi _{ab};\ \ \  \\
\mathcal{B}_{ab} &=&\left[ \hat{\nabla}^{c}\sigma
_{d(a}+\hat{\nabla}^{c}\varpi _{d(a}\right] \varepsilon _{b)ec}^{\ \ \ \ d}u^{e}; \\
\label{eq:Constraintphi} \hat{\nabla}^{b}\mathcal{E}_{ab} &=&
\frac{1}{2}\kappa \left[\hat{\nabla}^{b}\pi_{ab}+\frac{2}{3}\theta
q_{a}+\frac{2}{3}\hat{\nabla}_{a}\rho \right];\\
\hat{\nabla}^{b}\mathcal{B}_{ab} &=&\frac{1}{2}\kappa
\left[\hat{\nabla}_{c}q_{d}+(\rho +p)\varpi _{cd}\right]
\varepsilon _{ab}^{\ \ cd}u^{b},
\end{eqnarray}%
and five propagation equations,
\begin{eqnarray}
\label{eq:Raychaudhrui}
\dot{\theta}+\frac{1}{3}\theta^{2}-\hat{\nabla}^{a}A_{a}+\frac{\kappa}{2}(\rho +3p) &=& 0; \\
\label{eq:Propagationsigma} \dot{\sigma}_{ab}+\frac{2}{3}\theta
\sigma _{ab}-\hat{\nabla}_{\langle
a}A_{b\rangle }+\mathcal{E}_{ab}+\frac{1}{2}\kappa \pi _{ab} &=& 0; \\
\dot{\varpi}+\frac{2}{3}\theta \varpi -\hat{\nabla}_{[a}A_{b]} &=& 0; \\
\label{eq:Propagationphi} \frac{1}{2}\kappa\left[\dot{\pi}_{ab} +
\frac{1}{3}\theta\pi_{ab}\right] - \frac{1}{2}\kappa \left[(\rho
+p)\sigma_{ab}+\hat{\nabla}_{\langle a}q_{b\rangle}\right]\nonumber \\
-\left[\dot{\mathcal{E}}_{ab}+\theta\mathcal{E}_{ab}-\hat{\nabla}^{c}
\mathcal{B}_{d(a}\varepsilon_{b)ec}^{\ \ \ \ d}u^{e}\right] &=& 0;\\
\dot{\mathcal{B}}_{ab}+\theta
\mathcal{B}_{ab}+\hat{\nabla}^{c}\mathcal{E}_{d(a}\varepsilon _{b)ec}^{\ \ \ \ d}u^{e}\nonumber\\
+\frac{\kappa}{2}\hat{\nabla}^{c}\mathcal{\pi}_{d(a}\varepsilon_{b)ec}^{\
\ \ \ d}u^{e} &=& 0.
\end{eqnarray}%
Here, $\varepsilon _{abcd}$ is the covariant permutation tensor, $\mathcal{E}%
_{ab}$ and $\mathcal{B}_{ab}$ are respectively the electric and
magnetic
parts of the Weyl tensor $\mathcal{W}_{abcd}$, defined by $\mathcal{E}%
_{ab}=u^{c}u^{d}\mathcal{W}_{acbd}$ and $\mathcal{B}_{ab}=-\frac{1}{2}%
u^{c}u^{d}\varepsilon _{ac}^{\ \ ef}\mathcal{W}_{efbd}$. The angle
bracket means taking the trace-free part of a quantity.

Besides the above equations, it is useful to express the projected
Ricci scalar $\hat{R}$ into the hypersurfaces orthogonal to
$u^{a}$ as
\begin{eqnarray} \label{eq:SpatialRicciCurvature}
\hat{R} &\doteq& 2\kappa\rho - \frac{2}{3}\theta ^{2}.
\end{eqnarray}%
The spatial derivative of the projected Ricci scalar, $\eta _{a}\equiv \frac{%
1}{2}a\hat{\nabla}_{a}\hat{R}$, is then given as
\begin{eqnarray} \label{eq:Constrainteta}
\eta _{a} &=& \kappa \hat{\nabla}_{a}\rho - \frac{2a}{3}\theta\hat{\nabla}%
_{a}\theta,
\end{eqnarray}%
and its propagation equation by
\begin{eqnarray} \label{eq:Propagationeta}
\dot{\eta}_{a}+\frac{2\theta }{3}\eta _{a} &=& -\frac{2}{3}\theta a\hat{%
\nabla}_{a}\hat{\nabla}\cdot A -
a\kappa\hat{\nabla}_{a}\hat{\nabla}\cdot q.
\end{eqnarray}

Finally, there are the conservation equations for the
energy-momentum tensor:
\begin{eqnarray}
\label{eq:EnergyConservation} \dot{\rho}+(\rho +p)\theta +\hat{\nabla}^{a}q_{a} &=& 0, \\
\label{eq:HeatfluxEvolution} \dot{q}_{a}+\frac{4}{3}\theta
q_{a}+(\rho +p)A_{a}-\hat{\nabla}_{a}p+\hat{\nabla}^{b}\pi _{ab}
&=& 0.
\end{eqnarray}

As we are considering a spatially-flat universe, the spatial
curvature must vanish on large scales and so $\hat{R}=0$. Thus,
from Eq.~(\ref{eq:SpatialRicciCurvature}), we obtain
\begin{eqnarray}
\frac{1}{3}\theta ^{2}=\kappa\rho.
\end{eqnarray}
This is the Friedmann equation in standard general relativity, and
the other background equations (the Raychaudhuri equation and the
energy-conservation equation) can be obtained by taking the
zero-order parts of Eqs.~(\ref{eq:Raychaudhrui},
\ref{eq:EnergyConservation}), yielding:
\begin{eqnarray}
\dot{\theta}+\frac{1}{3}\theta ^{2}+\frac{\kappa }{2}(\rho +3p) &=& 0, \\
\label{eq:BackgroundEnergyConservation} \dot{\rho}+(\rho +p)\theta
&=& 0.
\end{eqnarray}

All through this paper we only consider scalar perturbation modes,
for which the vorticity $\varpi _{ab}$ and magnetic part of Weyl
tensor $\mathcal{B}_{ab}$ are at most of second order
\cite{GR3+1}, and so are neglected in our first-order analysis.

\subsection{Perturbation Equations in $k$-space}

\label{sect:appendixb}

For the perturbation analysis it is more convenient to work in the
$k$ space because we confine ourselves in the linear regime and
different $k$-modes decouple. Following \cite{GR3+1}, we shall
make the following harmonic expansions of our perturbation
variables
\begin{eqnarray} \label{eq:HarmonicExpansion}
\hat{\nabla}_{a}\rho = \sum_{k}\frac{k}{a}\mathcal{X}Q_{a}^{k}\ \
\ \ \hat{\nabla}_{a}p =
\sum_{k}\frac{k}{a}\mathcal{X}^{p}Q_{a}^{k}\nonumber\\ q_{a} =
\sum_{k}qQ_{a}^{k}\ \ \ \ \pi_{ab} = \sum_{k}\Pi Q_{ab}^{k}\ \ \ \
\nonumber\\
\hat{\nabla}_{a}\theta =
\sum_{k}\frac{k^{2}}{a^{2}}\mathcal{Z}Q_{a}^{k}\ \ \ \ \sigma_{ab}
= \sum_{k}\frac{k}{a}\sigma Q_{ab}^{k}\nonumber\\
\hat{\nabla}_{a}a = \sum_{k}khQ_{a}^{k}\ \ \ \ A_{a} =
\sum_{k}\frac{k}{a}AQ^{k}_{a}\nonumber\\
\mathcal{E}_{ab} = -\sum_{k}\frac{k^{2}}{a^{2}}\phi Q_{ab}^{k}\ \
\ \ \eta_{a} = \sum_{k}\frac{k^{3}}{a^{2}}\eta Q_{a}^{k}
\end{eqnarray}
in which $Q^{k}$ is the eigenfunction of the comoving spatial
Laplacian $a^{2}\hat{\nabla}^{2}$ satisfying
\begin{eqnarray}
\hat{\nabla}^{2}Q^{k} &=& \frac{k^{2}}{a^{2}}Q^{k}\nonumber
\end{eqnarray}
and $Q_{a}^{k},Q_{ab}^{k}$ are given by
$Q_{a}^{k}=\frac{a}{k}\hat{\nabla}
_{a}Q^{k},Q_{ab}^{k}=\frac{a}{k}\hat{\nabla}_{\langle
a}Q_{b\rangle }^{k}$.

In terms of the above harmonic expansion coefficients,
Eqs.~(\ref{eq:Constraintq}, \ref{eq:Constraintphi},
\ref{eq:Propagationsigma}, \ref{eq:Propagationphi},
\ref{eq:Constrainteta}, \ref{eq:Propagationeta}) can be rewritten
as \cite{GR3+1}
\begin{eqnarray}
\label{eq:Constraintq2} \frac{2}{3}k^{2}(\sigma - \mathcal{Z}) &=&
\kappa
qa^{2},\\
\label{eq:Constraintphi2} k^{3}\phi &=& -\frac{1}{2}\kappa
a^{2}\left[k(\Pi
+\mathcal{X})+3\mathcal{H}q\right],\\
\label{eq:Propagationsigma2} k(\sigma' + \mathcal{H}\sigma) &=&
k^{2}(\phi+A)
-\frac{1}{2}\kappa a^{2}\Pi,\\
\label{eq:Propagationphi2} k^{2}(\phi'+\mathcal{H}\phi) &=&
\frac{1}{2}\kappa a^{2}\left[k(\rho+p)\sigma+kq-\Pi'
-\mathcal{H}\Pi\right],\ \ \ \ \\
\label{eq:Constrainteta2} k^{2}\eta &=& \kappa\mathcal{X}a^{2} -
2k\mathcal{H}\mathcal{Z},\\
\label{eq:Propagationeta2} k\eta' &=& -\kappa qa^{2} -
2k\mathcal{H}A
\end{eqnarray}
where $\mathcal{H}=a^{\prime }/a=\frac{1}{3}a\theta $ and a prime
denotes the derivative with respect to the conformal time $\tau$
($ad\tau =dt$). Also, Eq.~(\ref{eq:HeatfluxEvolution}) and the
spatial derivative of Eq.~(\ref{eq:EnergyConservation}) become
\begin{eqnarray}
\label{eq:HeatfluxEvolution2} q' + 4\mathcal{H}q + (\rho+p)kA -
k\mathcal{X}^{p} +
\frac{2}{3}k\Pi &=& 0,\\
\label{eq:EnergyConservation2} \mathcal{X}' + 3h'(\rho+p) +
3\mathcal{H}(\mathcal{X}+\mathcal{X}^{p}) + kq &=& 0.
\end{eqnarray}

\subsection{The Main Equations}

Recall that we are treating the modifications to GR as an EDM term and so
include them in the (generalized) energy-momentum tensor $T_{ab}$ to
maintain the standard form of the Einstein equations. Thus, we can
distinguish its different components $\rho_{\mathrm{EDM}}, p_{\mathrm{EDM}},
q_{a,\mathrm{EDM}}, \pi_{ab,\mathrm{EDM}}$ and their conservation. Depending
on the specific model, the expressions for these components can be very
different, but their conservation equations will take the same form. In
particular, since the EDM has no coupling with standard model particles such
as photons and baryons, it satisfies a separated conservation equation,
Eqs.~(\ref{eq:HeatfluxEvolution2}, \ref{eq:EnergyConservation2}):
\begin{eqnarray}  \label{eq:EDMconserv1}
v^{\prime}_{\mathrm{EDM}}+\mathcal{H}v_{\mathrm{EDM}}+kA-k\Delta_{p} +\frac{2%
}{3}k\Delta _{\pi } &=& 0, \\
\label{eq:EDMconserv2}
\Delta^{\prime}_{\mathrm{EDM}}+k\mathcal{Z}-3\mathcal{H}A+3\mathcal{H}
\Delta_{p}+kv_{\mathrm{EDM}} &=& 0
\end{eqnarray}%
where we have defined the EDM peculiar velocity $v_{\mathrm{EDM}}\equiv q_{%
\mathrm{EDM}}/\rho _{\mathrm{EDM}}$, the density contrast $\Delta_{\mathrm{%
EDM}}\equiv \mathcal{X}_{\mathrm{EDM}}/\rho_{\mathrm{EDM}}$, and $%
\Delta_{p}\equiv \mathcal{X}_{\mathrm{EDM}}^{p}/\rho_{\mathrm{EDM}}$, $%
\Delta_{\pi }\equiv \Pi_{\mathrm{EDM}}/\rho_{\mathrm{EDM}}$ (c.f.~\S ~\ref%
{sect:appendixb}) for the EDM, and used the fact that $p_{\mathrm{EDM}}=0$
to reproduce the standard CDM background evolution. The prime here is the
derivative with respect to the conformal time and $\mathcal{H}=a^{\prime }/a$%
.

On the other hand, the spatial derivative of Eq.~(\ref{eq:Raychaudhrui})
gives the evolution equation for $\mathcal{Z}$ as
\begin{equation}  \label{eq:EvolveZ}
k^{\prime}\mathcal{Z}+k\mathcal{HZ}-kA+\frac{1}{2}\kappa (\mathcal{X}+3%
\mathcal{X}^{p})a^{2}=0
\end{equation}
and in this equation $\mathcal{X},\mathcal{X}^{p}$ are respectively the
density and pressure perturbations for \emph{all} the matter species,
including the EDM. For convenience we shall work in the frame where $A=0$.
In this case, if the universe is dominated by the EDM, then the three
equations above can be combined to eliminate $\mathcal{Z}$ and $v_{\mathrm{%
EDM}}$:
\begin{eqnarray}  \label{eq:EvolveDensityPerturb}
\Delta^{\prime\prime}_{\mathrm{EDM}}+\mathcal{H}\Delta_{\mathrm{EDM}}-\frac{1%
}{2}\kappa \rho_{\mathrm{EDM}}a^{2}\Delta_{\mathrm{EDM}}-\frac{2}{3}%
k^{2}\Delta_{\pi}  \notag \\
+3\mathcal{H}\Delta^{\prime}_{p}+\left[3\mathcal{H}^{\prime}+3\mathcal{H}%
^{2}-\frac{3}{2}\kappa \rho_{\mathrm{EDM}}a^{2}+k^{2}\right] \Delta_{p} &=&
0.\ \ \ \
\end{eqnarray}%
If other matter species could not be neglected, as is in the radiation
dominated era and early matter era, then we only need to correct the above
equation by adding to it $\frac{1}{2}\kappa \rho_{\mathrm{EDM}}a^{2}\Delta _{%
\mathrm{EDM}}$ another term with $\frac{1}{2}\kappa \rho _{b}a^{2}\Delta
_{b}+\kappa \rho_{r}a^{2}\Delta_{r}$ (where $\rho_{b},~\rho_{r}$ are the
baryon and radiation energy densities) which comes from the last term in the
left-hand side of Eq.~(\ref{eq:EvolveZ}) and these new terms do not affect
the qualitative features of our discussion (we will include them in the
numerical calculation). The Eq.~(\ref{eq:EvolveDensityPerturb}) tells us
that the evolution of the density perturbation in EDM is completely
controlled by its stress history: the EDM equation of state (EOS) $w_{%
\mathrm{EDM}}=p_{\mathrm{EDM}}/\rho_{\mathrm{EDM}}$ controls the background
expansion and is set to zero here; the other two stress variables $%
\Delta_{p},~\Delta_{\pi}$ then drive the evolution of $\Delta_{\mathrm{EDM}}$
through Eq.~(\ref{eq:EvolveDensityPerturb}). Note that the stresses of the
EDM are external functions determined by microphysics (for particle dark
matter) or a particular modified gravity theory, and must be specified to
close the system of Einstein equations and conservation equations. In the
special case where $\Delta_{p}=\Delta_{\pi}=0$, we reduce to the CDM model
for which $\Delta_{\mathrm{CDM}}\propto a$ in the matter era; but in general
there will be deviations from this growing solution.

Next, we look at the evolution of the gravitational potential, $\phi $ (see
\S ~\ref{sect:appendixa}, \ref{sect:appendixb}). By manipulating Eqs.~(\ref%
{eq:Constraintphi2}, \ref{eq:Propagationsigma2}, \ref{eq:Propagationphi2}, %
\ref{eq:HeatfluxEvolution2}), and working again in the $A=0$ frame, we can
eliminate the terms involving $q$ and obtain the following evolution
equation
\begin{widetext}
\begin{eqnarray} \label{eq:EvolvePotential}
&& \phi'' + 3\mathcal{H}\left(1+\frac{p'}{\rho'}\right)\phi' +
\left[2\mathcal{H}' + \mathcal{H}^{2}\left(1 +
3\frac{p'}{\rho'}\right)\right]\phi +
k^{2}\frac{p'}{\rho'}\phi\nonumber\\
&=& \frac{1}{2}\kappa\rho a^{2}\left(\Delta_{p} -
\frac{p'}{\rho'}\Delta\right) - \frac{1}{2k^{2}}\kappa\rho
a^{2}\Delta''_{\pi} - \frac{1}{2k^{2}}\kappa\rho
a^{2}\left[\left(5+3\frac{p'}{\rho'}\right)\mathcal{H} +
2\frac{\rho'}{\rho}\right]\Delta'_{\pi}\nonumber\\
&& - \frac{1}{2k^{2}}\kappa\rho a^{2}\left[\left(5 +
3\frac{p'}{\rho'}\right)\left(\mathcal{H} +
\frac{\rho'}{\rho}\right)\mathcal{H} + \frac{\rho''}{\rho} +
\left(\frac{2}{3} +
\frac{p'}{\rho'}\right)k^{2}\right]\Delta_{\pi}
\end{eqnarray}
\end{widetext}where $\rho ,p$ are the total energy density and total
pressure respectively and $\rho \equiv \rho _{\mathrm{EDM}}$, with $p=0$ for
an EDM dominated Universe. The term $\frac{1}{2}\kappa\rho a^{2}\left(
\Delta_{p}-\frac{p^{\prime}}{\rho^{\prime}}\Delta \right) $ is zero for
radiation and baryons, but in general could be nonzero for the EDM. Again,
we see that the stress history of the EDM completely specifies the evolution
of $\phi$. As we shall see in what follows, the dark matter component
influences the CMB power spectrum through the gravitational potential, and
so the stress history of the EDM is of crucial importance for the CMB.

In summary, we have seen that the properties of the (isotropic and
anisotropic) stresses of the EDM determine the evolution of the matter
density perturbation and the gravitational potential, and thereby determine
the predicted matter power spectrum (through the former) and the CMB
spectrum (through the latter) \cite{Hu1999}. As we discussed in \S ~\ref%
{sect:Introduction}, the predicted matter power spectrum of a theory records
the matter energy density perturbation at a specific time -- today, and it
is easier to make it consistent with observations, as evident from the
studies of TeVeS and $f(K)$ theories. Thus, in this work we shall start from
an EDM theory which is constructed already to predict an acceptable matter
power spectrum; that is, we fix the matter energy density perturbation today
and calculate the influence of different $\delta_{m}$ evolution paths on the
CMB spectrum. In this way we can reduce our model space to one that is of
realistic interests.

\section{The Dark Matter Effects on CMB}

\label{sect:CMB}

It is helpful to have a brief review of the CMB physics and how it is
affected by the dark matter before we go into the numerics to show the
effects of EDM stresses on the CMB. Here we will just present a minimal
description of this topic, for more details see the reviews \cite{Hu1997,
Hu2002, Hu2008, Dodelsonbook}.

The primary CMB spectrum is determined by inhomogeneities in the CMB photon
temperature at the time of recombination. Prior to the recombination,
photons couple tightly by Thomson scattering with electrons which themselves
couple to baryons via Coulomb interaction, thus to a first approximation
photons and baryons combine into a single fluid. Dark matter, on the other
hand, does not couple electromagnetically, but only interacts through
gravitational effects and so contributes to the gravitational potential in
which the baryon-photon fluid moves.

When the density perturbation in the photons grows, hot (cold) spots appear
where the local photon density is higher (lower) than average. The higher
photon density means higher pressure, which will tend to counteract the
growth of local photon density. This will lead to acoustic oscillations of
the gauge invariant photon temperature perturbation $\Theta $, which is also
a characterization of the photon number density perturbation. In reality
this picture becomes a bit more complicated because of the interplay between
baryons and dark matter: baryons couple tightly to photons, and move with
them so that the inertia of the fluid is increased, while the gravitational
potential produced by dark matter drives the oscillations. More
specifically, if we use the multipole decomposition
\begin{eqnarray}
\Theta &=& \sum_{\ell}(-i)^{\ell}\Theta_{\ell}P_{\ell}(\mu),  \notag
\end{eqnarray}
where $P_{\ell }$ is the Legendre function and $k\mu =\mathbf{k}\cdot
\mathbf{\gamma }$ with $\mathbf{\gamma }$ being the direction of photon
momentum, then in the tight-coupling limit the monopole $\Theta _{0}$
satisfies a driven-oscillator equation \cite{Hu1995}:
\begin{eqnarray}  \label{eq:DrivenOscillator}
\Theta^{\prime\prime}_{0}+\frac{R}{1+R}\mathcal{H}\Theta^{\prime}_{0}
+k^{2}c_{s}^{2}\Theta_{0} &=& F,
\end{eqnarray}
in which
\begin{eqnarray}
F &=& -\Phi^{\prime\prime}-\frac{R}{1+R}\mathcal{H}\Phi^{\prime}-\frac{1}{3}
k^{2}\Psi,
\end{eqnarray}
where
\begin{equation*}
R\equiv 3\rho _{b}/4\rho _{\gamma }, \qquad c_{s}^{2}\equiv 1/3(1+R)
\end{equation*}
and
\begin{equation*}
\Psi =\phi -\frac{1}{2}\frac{a^{2}}{k^{2}}\kappa \Pi,\qquad \Phi =-\phi -%
\frac{1}{2}\frac{a^{2}}{k^{2}}\kappa \Pi
\end{equation*}
are, respectively, the (frame-independent) Newtonian potential and curvature
perturbations. The dipole moment $\Theta _{1}$, which equals the peculiar
velocity of baryons thanks to the tight coupling, satisfies $k\Theta _{1}+3%
\mathcal{H}\Theta _{0}^{\prime }+3\Phi ^{\prime }=0$, and all higher moments
$\Theta _{\ell }~(\ell \geq 2)$ vanish because the frequent scattering makes
the photon distribution isotropic in the electron rest frame.

Thus, on large scales the photon temperature perturbation displays a pattern
of driven and damped oscillation. On scales smaller than the photon mean
free path, which itself grows in time, however, the tight coupling
approximation is no longer perfect and quadrupole moments of the temperature
perturbation needs to be taken into account. This introduces a dissipation
term into the oscillator equation above, which damps the oscillations.

At the time of recombination, the number density of free electrons drops
suddenly and there ceases to be coupling between baryons and photons. The
CMB photons then free stream towards us. This free-streaming solution is
given by \cite{Hu1995}
\begin{eqnarray}  \label{eq:FreeStreaming}
\Theta_{\ell}(\tau_{0}) &\approx& (\Theta_{0}+\Psi)(\tau_{\ast})(2\ell
+1)j_{\ell}(k\Delta \tau_{\ast})  \notag \\
&& +\Theta_{1}(\tau_{\ast})\left[\ell j_{\ell-1}(k\Delta
\tau_{\ast})-(\ell+1)j_{\ell +1}(k\Delta\tau_{\ast})\right]  \notag \\
&&+(2\ell+1)\int_{\tau_{\ast}}^{\tau_{0}}(\Psi^{\prime}-\Phi^{\prime})j_{%
\ell}[k(\tau_{0}-\tau)]d\tau
\end{eqnarray}
where $\tau _{0},\tau _{\ast }$ are the conformal times today and at
recombination, $\Delta \tau _{\ast }=\tau _{0}-\tau _{\ast }$ and $j_{\ell
}(x)$ is the spherical Bessel function of order $\ell $. The terms $\Theta
_{0}+\Psi $ and $\Theta _{1}$ are respectively the monopole and dipole
moments of the CMB temperature field at $\tau _{\ast },$ and can be
calculated using the equations above (a $\Psi $ is added to $\Theta _{0},$
which accounts for the redshift in the photon energy, and thus temperature,
when it climbs out of the Newtonian potential). Finally, the CMB
temperature-temperature spectrum is defined as
\begin{eqnarray}  \label{eq:Cl}
\frac{2\ell +1}{4\pi }C_{\ell }=\frac{V}{2\pi ^{2}}\int \frac{k^{3}|\Theta
_{\ell }(\tau _{0},k)|^{2}}{2\ell +1}d\ln k.
\end{eqnarray}

These are the leading-order effects in the CMB physics, and we could see how
they affect the CMB power spectrum. Because $j_{\ell }(x)$ peaks strongly at
$\ell \sim x$, so perturbations on very large scales (small $k$) mainly
affect the low-$\ell $ spectrum. If the Universe is completely dominated by
matter at $\tau _{\ast }$ then the first term in Eq.~(\ref{eq:FreeStreaming}%
) is given by $(\Theta _{0}+\Psi )(\tau _{\ast })\approx \Psi (\tau _{\ast
})/3,$ and accounts for the ordinary Sachs-Wolfe (SW) effect; meanwhile, if
the potential $\Psi -\Phi =2\phi $ decays between $\tau _{\ast }$ and $\tau
_{0}$, then the integration in Eq.~(\ref{eq:FreeStreaming}) will also make a
significant contribution as photons travel in and out of many time-dependent
potentials along the line of sight, and this is the integrated Sachs-Wolfe
(ISW) effect.

Going to higher $\ell $ one can see the peak structures of the CMB spectrum.
The peaks appear since at $\tau _{\ast }$, when the oscillating pattern of $%
\Theta_{0}$ [c.f.~Eq.~(\ref{eq:DrivenOscillator})] freezes, $\Theta _{0}$
might be just at its extrema for some scales ($k$); and these extrema are
then converted to extrema of $C_{\ell }$ through Eqs.~(\ref{eq:FreeStreaming}%
, \ref{eq:Cl}). Since what appears in Eq.~(\ref{eq:Cl}) is $|\Theta _{\ell
}|^{2}$, both the maxima and minima of $(\Theta _{0}+\Psi )(\tau _{\ast })$
will appear as peaks in $C_{\ell }$. If there are no baryons and the
potential $\Psi $ is constant, then the even and odd peaks should be of the
same amplitude. The inclusion of baryons effectively increases the inertia
of the baryon-photon fluid and displaces the balance point of $\Theta
_{0}+\Psi $, and as a result, after taking $|\cdots |^{2},$ the odd and even
peaks appear to have different heights. At the same time, from Eqs.~(\ref%
{eq:FreeStreaming}, \ref{eq:Cl}), we see that the dipole $\Theta _{1}(\tau
_{\ast })$ also contributes to $C_{\ell }$ through modulation. But Eq.~(\ref%
{eq:FreeStreaming}) shows that its power is more broadly distributed and so
its contribution is significantly smaller than that of the monopole. Where
the monopole vanishes the dipole becomes important and this is why the
troughs of $C_{\ell }$ are of nonzero amplitude.

If one goes to still higher $\ell ,$ there are two effects. First, the
higher $\ell $ moments mainly receive contributions from the small-scale
(large $k$) perturbations, which began oscillating already during the
radiation-dominated era (not much earlier than $\tau _{\ast }$). Since in
the radiation era the potential $\Phi $ decays, this decay will drive the
monopole oscillation through the source term in Eq.~(\ref%
{eq:DrivenOscillator}) and lead to an increase in the oscillation
amplitudes. This explains why in some models the third peak is higher than
the second. Second, as discussed above, on very small scales there is severe
damping of the oscillations because of photon diffusion. The combination of
these two effects causes the CMB power in $\ell $s higher than the third
peak to be significantly damped.

We can now highlight the key places where the dark-matter effects enter CMB
physics through the gravitational potential it produces and contrast the
situation with that when EDM is employed as a substitute. First, if CDM is
replaced by EDM, then in the analysis of \S ~\ref{sect:Model} the time
evolutions of $\phi ,$ and thus of $\Phi $ and $\Psi $ are modified. If the
modification only becomes significant after $\tau _{\ast }$, the ISW effect
could be different from standard CDM (where it effectively vanishes). If it
differs from standard CDM before $\tau _{\ast }$ then the SW effect will be
altered as well. Second, the deviation of $\Phi $ and $\Psi $ from their
values in standard CDM before $\tau _{\ast }$ can change the zero point of
the monopole oscillation through baryon loading, thereby modifying the
relative heights of the odd and even peaks. Third, the modification of $\Phi
$ and $\Psi $ also changes the driving force in Eq.~(\ref%
{eq:DrivenOscillator}), and results in different amplitudes for the $C_{\ell
}$ at high $\ell $ . In \S ~\ref{sect:Numerics} we shall give some numerical
examples showing these effects explicitly in our EDM model.

\section{Numerical Examples}

\label{sect:Numerics}

In this section we turn to some numerical examples to illustrate the
qualitative analysis of \S ~\ref{sect:CMB}. As mentioned in \S ~\ref%
{sect:Model}, we choose to fix the endpoint of the evolution of the matter
energy density perturbation. This gives us a freedom to parametrize the
evolution of $\Delta _{\mathrm{EDM}}(a)$, which is generally different from
that of $\Delta _{\mathrm{CDM}}(a)\propto a$ (in matter regime). Once the $%
\Delta _{\mathrm{EDM}}$ parametrization is given (or in other words the
model is set up), Eq.~(\ref{eq:EvolveDensityPerturb}) becomes an evolution
equation for $\Delta _{p}$ and $\Delta _{\pi }$. This is insufficient to
solve for $\Delta _{p}$ and $\Delta _{\pi }$, so in the following numerical
calculations we consider three cases: (1) $\Delta _{\pi }=0$, (2) $\Delta
_{p}=0,$ and (3) a more realistic case where both $\Delta _{p}$ and $\Delta
_{\pi }$ are nonzero. Since any scale dependence of $\Delta _{\mathrm{EDM}}$
will alter the shape of the matter power spectrum, and thus incur stringent
constraints, we parametrize $\Delta _{\mathrm{EDM}}(a)$ so that it is
independent of $k$.

\subsection{The Case of $\Delta_{\protect\pi} = 0$}

\label{subsect:case1}

\footnotetext[1]{Here we include the effects of baryons and
radiation explicitly.}

When $\Delta_{\pi}=0$ and $\Delta_{\mathrm{EDM}}$ is specified, Eq.~(\ref%
{eq:EvolveDensityPerturb}) becomes a first-order evolution equation for $%
\Delta_{p}$
\begin{eqnarray}  \label{eq:case1Delta_p}
3\mathcal{H}\Delta^{\prime}_{p}+\left[3\mathcal{H}^{\prime}+3\mathcal{H}^{2}-%
\frac{3}{2}\kappa \rho_{\mathrm{EDM}}a^{2}+k^{2}\right]\Delta_{p} &=& S,\ \
\ \
\end{eqnarray}
where the source term $S$ is given by \footnotemark[1]
\begin{eqnarray}
S &=& -\Delta^{\prime\prime}_{\mathrm{EDM}}-\mathcal{H}\Delta^{\prime}_{%
\mathrm{EDM}}+\frac{1}{2}\kappa \rho_{\mathrm{EDM}}a^{2}\Delta_{\mathrm{EDM}}
\notag \\
&& +\frac{1}{2}\kappa \rho_{b}a^{2}\Delta_{b}+\kappa \rho_{r}a^{2}\Delta_{r}.
\end{eqnarray}
The quantity $\Delta_{p}$ does not appear directly in the expression for
gravitational potential [c.f.~Eqs.~(\ref{eq:Constraintphi2}, \ref%
{eq:Propagationphi2})], however it does affect the potential $\phi $ \emph{%
indirectly} through the evolution of $v_{\mathrm{EDM}}$ Eq.~(\ref%
{eq:EDMconserv1}). Consequently, there are now two new variables ($%
\Delta_{p} $ and $v_{\mathrm{EDM}}$) to evolve in the model and we need to
specify their initial conditions.

Here, we adopt the simplest and most direct approach, namely to
assume that the EDM evolves as CDM prior to some initial time
$\tau _{i}$, when the scale factor is $a_{i}$ and the dark matter
density perturbation $\Delta _{i} $, and for $a>a_{i}$ the dark
matter density perturbation begins to evolve as $\Delta
_{\mathrm{EDM}}(a)$. This means that for $a<a_{i}$ the variables
$\Delta _{p}$ and $v_{\mathrm{EDM}}$ \footnotemark[2] will remain
zero: this provides the desired initial conditions. This simple
approach captures most of the interesting features about EDM.
However, by assuming that at early times EDM is just CDM, we
cannot account for more complicated issues such as the primordial
power spectrum, and we will comment on this in the concluding
section.

\footnotetext[2]{Recall that we work in the $A=0$ frame, in which
$v_{\mathrm{CDM}}=0$.}

\begin{figure}[tbp]
\centering \includegraphics[scale=0.9] {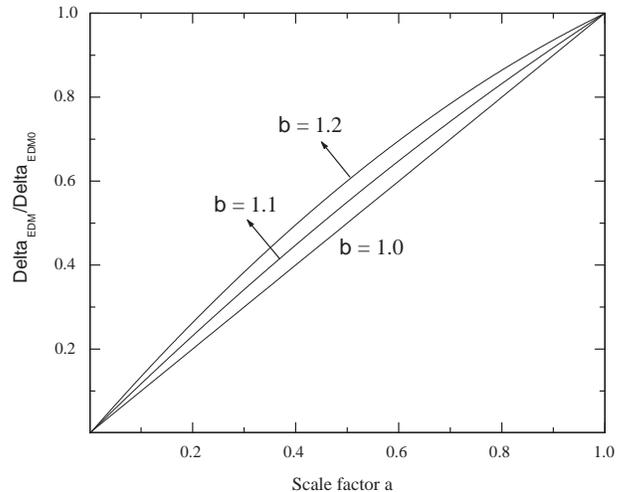}
\caption{Some example parameterizations for the evolution of $\Delta_{%
\mathrm{EDM}}$, normalized to its current value $\Delta_{\mathrm{EDM}0}$, as
described in Eq.~(\protect\ref{eq:EDMparametrization}). The values of the
parameter $b$ are indicated; the other parameters are $a_{i} = 0.005$ and $%
a_{f} = 1$.}
\label{fig:Figure1}
\end{figure}

Now we could turn to a specific example of
$\Delta_{\mathrm{EDM}}(a)$
parametrization. We let $\Delta _{\mathrm{EDM}}(a)$ equal $\Delta_{\mathrm{%
CDM}}(a)$ in the standard CDM model at times $a_{i}$ and $a_{f}$, and assume
that the deviation from standard CDM occurs between $a_{i}$ and $a_{f}$.
Here, $a_{f}\leq a_{0}$ and $a_{0}=1$ is the current time. In order to
characterize the deviation from standard CDM between $a_{i}$ and $a_{f}$, we
introduce a parameter $b$ to denote the ratio of $\Delta _{\mathrm{EDM}}$ to
$\Delta _{\mathrm{CDM}}$ at the mean time $a=(a_{i}+a_{f})/2$. Since for
standard CDM we have $\Delta _{\mathrm{CDM}}(a)\propto a$, the parametrized $%
\Delta _{\mathrm{EDM}}$ is simply a parabola which passes through three
points $(a_{i},~\Delta _{i})$, $(a_{f},~\Delta _{i}a_{f}/a_{i})$ and $(\frac{%
a_{i}+a_{f}}{2},~b\frac{\Delta _{i}}{a_{i}}\frac{a_{i}+a_{f}}{2})$ between
times when the scale factor lies between $a_{i}$ and $a_{f}$:
\begin{equation}  \label{eq:EDMparametrization}
\Delta _{\mathrm{EDM}}(a)=\frac{\Delta _{i}}{a_{i}}a-2\frac{\Delta _{i}}{%
a_{i}}\frac{a_{f}+a_{i}}{(a_{f}-a_{i})^{2}}(b-1)(a-a_{i})(a-a_{f})\ \
\end{equation}
for $a\in \lbrack a_{i},~a_{f}]$. Some examples of $\Delta _{\mathrm{EDM}%
}(a) $ with different choices of $b$ are shown in Fig.~\ref{fig:Figure1}. It
clearly shows that $b$ characterizes the deviation from a standard CDM
evolution ($b=0$). Note that the parametrization Eq.~(\ref%
{eq:EDMparametrization}) is just an example for illustration, and specific
modified gravity models like TeVeS and $f(K)$ theory may lead to different
parametrization which however can similarly be tested.

\begin{figure}[tbp]
\centering \includegraphics[scale=0.9] {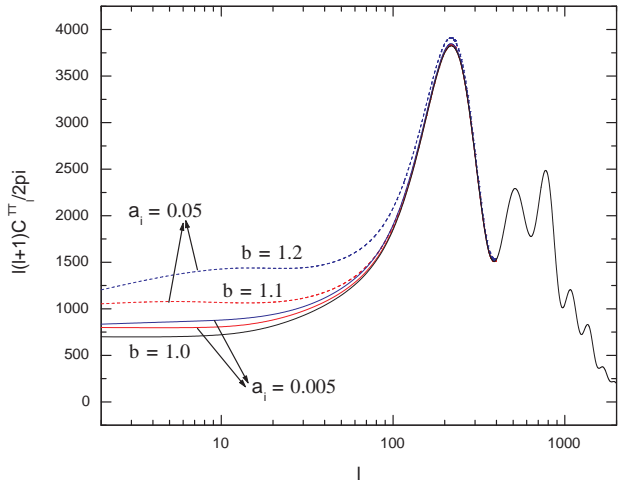}
\caption{(color online) The CMB power spectrum for our $\Delta _{\protect\pi %
}=0$ model with $\Delta _{\mathrm{EDM}}(a)$ parameterized as in Eq.~(\protect
\ref{eq:EDMparametrization}). Three values of $b$ are adopted: $b=1.2$ (blue
curves), $b=1.1$ (red curves) and $b=1.0$ (black curve). The dashed curves
are the cases where $a_{i}=0.005$ and the solid coloured ones $a_{i}=0.05$.
For all the curves we set $a_{f}=1.0$.}
\label{fig:Figure2}
\end{figure}

Then with the parametrization of $\Delta _{\mathrm{EDM}}(a)$ in Eq.~(\ref%
{eq:EDMparametrization}), we could numerically evolve the relevant
perturbation equations and see the changes in the CMB power spectrum. First,
we assume $\tau_{i} > \tau_{\ast}$. In this situation the CMB physics prior
to last scattering is unaffected and so from Eq.~(\ref{eq:FreeStreaming}) we
see that only the integration is modified. This is because in the standard
CDM model the gravitational potential, $\phi = (\Psi -\Phi )/2$, remains
constant during the matter epoch, and the integrand in Eq.~(\ref%
{eq:FreeStreaming}) vanishes. For the present EDM case, however, Eq.~(\ref%
{eq:EvolvePotential}) dictates that $\phi^{\prime}\neq 0$ even in the matter
era; consequently, there will be a significant ISW effect to boost the low-$%
\ell$ CMB power. As shown in Fig.~\ref{fig:Figure2}, the earlier the EDM
evolution deviates from that of CDM (\emph{i.e.}, the smaller $a_{i}$ is),
the earlier $\phi $ begins to evolve and the larger the cumulative ISW
effect is.

\begin{figure}[tbp]
\centering \includegraphics[scale=0.9] {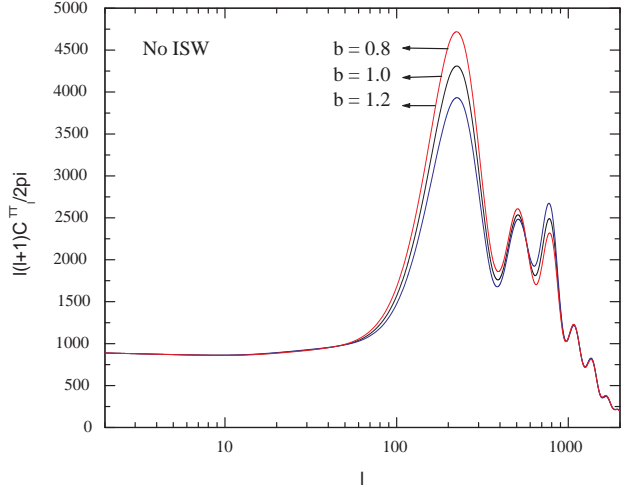}
\caption{(color online) The primary CMB spectrum with the ISW contribution
removed, for the parameters $a_{i} = 0.0002$, $a_{f} = 0.002$ and $b = 1.2$
(blue curve), $1.0$ (black curve) and $0.8$ (red curve) respectively.}
\label{fig:Figure4}
\end{figure}

\footnotetext[1]{%
Note that in reality the universe is not completely matter dominated between
$a=0.0002$ and $a=0.002$, and as a result the parametrization Eq.~(\ref%
{eq:EDMparametrization}), with $b =1$, is not exactly the same as $\Delta_{%
\mathrm{CDM}}(a)$. The qualitative features, however, are not affected by
this small deviation.}

Next, consider the case where the deviation from CDM evolution begins
earlier than last scattering. In this case the evolution of $\Delta_{\mathrm{%
EDM}}$ and thus the potential $\phi $ is changed before last scattering, and
correspondingly the first two terms in Eq.~(\ref{eq:FreeStreaming}) are
modified as well. In Fig.~\ref{fig:Figure4} we have displayed the primary
CMB power spectrum for the $\Delta _{\mathrm{EDM}}(a)$ parametrization Eq.~(%
\ref{eq:EDMparametrization}) with parameters $a_{i}=0.0002$, $a_{f}=0.002$
\footnotemark[1] , $b =1.2,1.0,0.8$ and the ISW effect switched off for
simplicity. A larger value of $b$ implies a larger EDM density perturbation
(for the scales relevant to the first acoustic peak) before the last
scattering and therefore a deeper Newtonian potential $\Psi$. This means
that the CMB photons experience larger redshifts when climbing out of the
potential after last scattering, and the effective temperature $%
(\Theta_{0}+\Psi )(\tau_{\ast })$ is lower, leading to a suppressed first
CMB acoustic peak. When $b$ is smaller the opposite effect occurs (c.f.~Fig.~%
\ref{fig:Figure4}). At the higher-$\ell$ peaks this effect is less
significant because the potential for the relevant (smaller) scales has
already decayed. However, as discussed in \S ~\ref{sect:CMB}, the change the
potential prior to last scattering also modifies the equilibrium point of
the monopole oscillation and alters the relative heights of odd and even CMB
peaks. If $\Psi$ is a constant, the difference between odd and even peaks in
$|\Theta_{0}+\Psi |$ is proportional to $|\Psi|$ \cite{Hu2002}, so
increasing $b$ amplifies this difference by increasing $\Psi$, and as shown
in Fig.~\ref{fig:Figure4}, the third peak becomes higher and the second peak
becomes lower for larger values of $b $ and go oppositely for smaller $b $.
In this figure the fourth peak and onwards are not affected significantly
because our parameters are chosen conservatively. There however \emph{could}
be alternative dark matter models where the higher peaks also deviate from
the CDM predictions, for an example see Fig.~4 (upper panel) of \cite%
{TeVeSperturbation}.

The CMB power spectrum has been measured to high precision by many
experiments (see for example \cite{WMAP3yr, ACBAR}) and will be further
improved in the future, so it could be used to constrain the EDM model here.
For higher $\ell $s the CMB data could be used directly. For lower $\ell $s
its usability is limited by the cosmic variance. If the deviation is
significant (such as those in Fig.~\ref{fig:Figure2}), then we could use CMB
data alone to constrain the model, as in \cite{Caldwell2007}. Otherwise, we
could cross correlate the observed ISW with the matter density perturbation
observable \cite{Hu2004}; because both are modified in the EDM model, we
should expect different correlations from those created by CDM. Actually,
this technique has been applied to TeVeS \cite{Schmidt2007} but may not
serve as a perfect EDM discriminator because there will be contaminations
from the dark energy component in the galaxy-CMB correlation. Meanwhile, the
general EDM model has different gravitational potential, matter density
perturbation, as well as a possible different redshift distribution of
lensing galaxies, and these will also change the weak lensing spectrum (see
\cite{Schafer2008} for an application to one modified dark matter model).
Finally, the different evolution history of $\Delta _{\mathrm{EDM}}(a)$ may
also have implications for the formation of nonlinear structure. These
further possibilities are beyond the scope of this work and will be further
pursued elsewhere.

\begin{figure}[tbp]
\centering \includegraphics[scale=0.9] {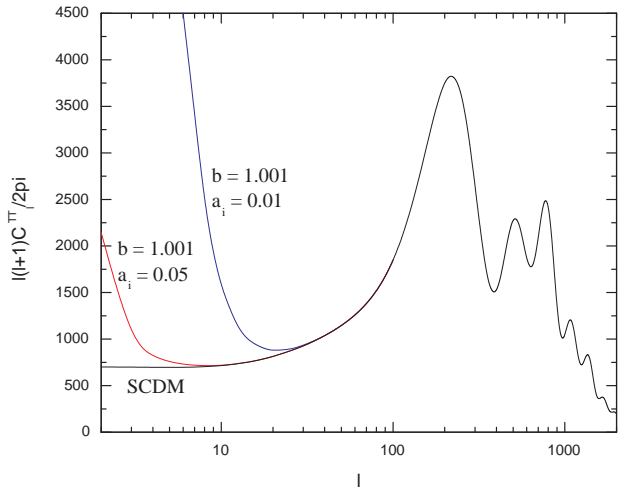}
\caption{(color online) The CMB spectrum of the model $\Delta_{p} = 0$ with
the $\Delta_{\mathrm{EDM}}(a)$ parametrization given in Eq.~(\protect\ref%
{eq:EDMparametrization}) and parameters $a_{i} = 0.01$ (blue curve) and $%
a_{i} = 0.05$ (red curve) respectively. The black curve is the SCDM model.
The other parameters are $b = 1.001$ and $a_{f} = 1.0$.}
\label{fig:Figure5}
\end{figure}

\subsection{The Case of $\Delta_{p} = 0$}

\label{subsect:case2}

In the case of $\Delta _{p}=0$, Eq.~(\ref{eq:EvolveDensityPerturb}) simply
becomes an algebraic equation for $\Delta _{\pi }$:
\begin{eqnarray}  \label{eq:case2Delta_pi}
\frac{2}{3}k^{2}\Delta_{\pi} &=& \Delta^{\prime\prime}_{\mathrm{EDM}}+
\mathcal{H}\Delta^{\prime}_{\mathrm{EDM}}-\frac{1}{2} \kappa\rho_{\mathrm{EDM%
}}a^{2}\Delta_{\mathrm{EDM}}  \notag \\
&&-\frac{1}{2}\kappa\rho_{b}a^{2}\Delta_{b}-\kappa\rho_{r}a^{2}\Delta_{r}.
\end{eqnarray}
For the parametrization of $\Delta_{\mathrm{EDM}}(a)$ we will still use Eq.~(%
\ref{eq:EDMparametrization}). In this case, because Eq.~(\ref%
{eq:case2Delta_pi}) is algebraic, the only variable to propagate in time is $%
v_{\mathrm{EDM}}$ [c.f.~Eq.~(\ref{eq:EDMconserv1})], and as in \S ~\ref%
{subsect:case1} we could take $v_{\mathrm{EDM}}=0$ prior to $a_{i}$ and then
evolve it using Eq.~(\ref{eq:EDMconserv1}) for $a>a_{i}$.

In Fig.~\ref{fig:Figure5} we have plotted the CMB spectrum for such a model
with $b =1.001$ and different choices of $a_{i}$, from which we can see that
the low-$\ell $ CMB power is very sensitive to both $b $ and $a_{i}$ . The
reason is that, although the right-hand side of Eq.~(\ref{eq:case2Delta_pi})
is scale-independent thanks to our parametrization Eq.~(\ref%
{eq:EDMparametrization}), its left-hand side \emph{is} scale-dependent
through the $k^{2}$ factor. Consequently, on large scales (small $k$) the
EDM anisotropic stress $\Delta_{\pi }$ could be very large and so the
gravitational potential $\phi$ is significantly different than that in
standard CDM[c.f.~Eq.~(\ref{eq:Constraintphi2})]. The late ISW effect is
then modified, which enhances the low-$\ell $ CMB power. On smaller scales, $%
\Delta _{\pi }$ is suppressed by $k^{-2}$ and its effects soon becomes
negligible, explaining why the high-$\ell $ CMB power is not influenced.
Also, the smaller $a_{i}$ is, so the earlier the CMB evolution of the $%
\Delta _{p}=0$ EDM model deviates from the standard CDM result. Note that,
in the case of standard CDM, the right hand side of Eq.~(\ref%
{eq:case2Delta_pi}) vanishes identically, and so there is no influence on
the last ISW.

It is then clear that the $\Delta _{p}=0$ EDM model with a general
scale-independent parametrization of $\Delta_{\mathrm{EDM}}(a)$ is
problematic and already stringently constrained. A possible way out of this
trouble is to drop the scale-independence of $\Delta_{\mathrm{EDM}}(a)$. We
have numerically checked that if on large scales $\Delta_{\mathrm{EDM}}$
grows as $\Delta_{\mathrm{CDM}}$, then the low-$\ell$ boosts of the CMB
power as shown in Fig.~\ref{fig:Figure5} disappear and one recaptures the
standard CDM results of ISW effect. Another possibility is to have both $%
\Delta_{\pi}$ and $\Delta_{\mathrm{EDM}}(a)$ scale-independent and also
include $\Delta_{p}$; this will be considered in \S ~\ref{subsect:case3}.

\begin{figure}[tbp]
\centering \includegraphics[scale=0.9] {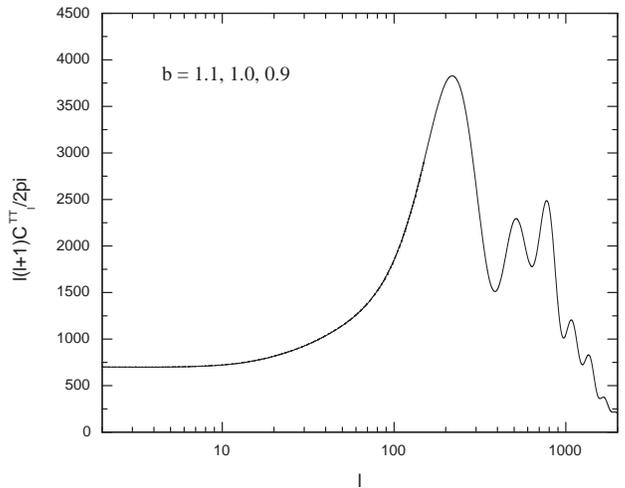}
\caption{The CMB power spectrum for the model in which $\Delta_{\mathrm{EDM}%
}(a)$ is parameterized as Eq.~(\protect\ref{eq:EDMparametrization}), both $%
\Delta_{p}$ and $\Delta_{\protect\pi}$ are nonzero and satisfy a relation as
described in the text. The solid, dashed and dotted curves represent the
cases of $b=1.0, 1.1, 0.9$ respectively, and they cannot be distinguished in
the figure because they experience identical gravitational potential
evolution. The other parameters are $a_{i} = 0.005$ and $a_{f} = 1$.}
\label{fig:Figure6}
\end{figure}

\begin{figure}[tbp]
\centering \includegraphics[scale=0.9] {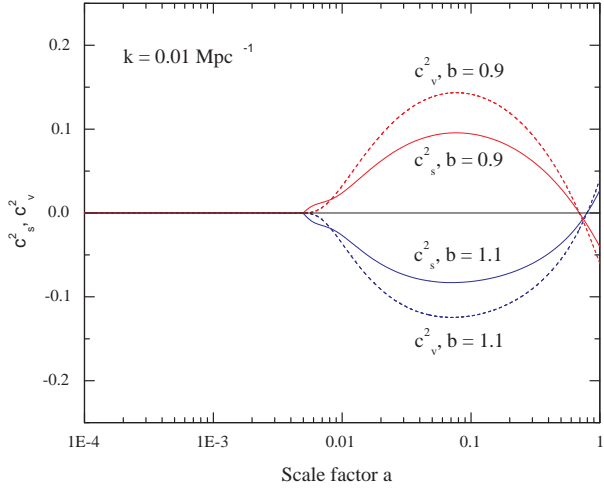}
\caption{(color online) The quantities $c^{2}_{s} \equiv \Delta_{p}/\Delta_{%
\mathrm{EDM}}$ (solid curves) and $c^{2}_{v} \equiv \Delta_{\protect\pi%
}/\Delta_{\mathrm{EDM}}$ (dashed curves) as functions of $a$. The black line
is for SCDM ($c^{2}_{s} = c^{2}_{v} = 0$), and the red/blue lines are for
the model described in the text with $b = 0.9, 1.1$ respectively. All curves
are for the scale $k = 0.01~\mathrm{Mpc}^{-1}$.}
\label{fig:Figure7}
\end{figure}

\subsection{General $\Delta_{p}$ and $\Delta_{\protect\pi}$}

\label{subsect:case3}

If the EDM arises from modifications to standard general relativity, then in
general neither $\Delta_{p}$ nor $\Delta_{\pi}$ would be exactly zero. In
this case there are many more possibilities because $\Delta _{p}$ and $%
\Delta_{\pi}$ cannot be uniquely solved from Eq.~(\ref%
{eq:EvolveDensityPerturb}). Of course, the evolution of $\phi$ might also be
modified normally and this could be used to constrain $\Delta_{p}$ and $%
\Delta_{\pi}$. Indeed, we could use the freedom to choose $\Delta_{p}$ and $%
\Delta _{\pi }$ so that the evolution of $\phi$ [c.f.~Eq.~(\ref%
{eq:EvolvePotential})] is exactly (or nearly) the same as that in standard
CDM. Let us consider such an example now.

In order that the gravitational potential evolves in the same way as in
standard CDM, we require the contribution on the right-hand side of Eq.~(\ref%
{eq:EvolvePotential}) from the EDM to vanish, which leads to the following
evolution equation for $\Delta _{\pi }$
\begin{eqnarray}  \label{eq:case3Delta_pi}
\Delta^{\prime \prime }_{\pi }-\frac{3\rho_{m}}{4\rho_{r}+3\rho_{m}}
\mathcal{H}\Delta^{\prime }_{\pi} \\
+\left[\frac{4\rho_{r}+2\rho_{m}}{4\rho_{r}+3\rho_{m}}k^{2}-3\mathcal{H}%
^{\prime }-\frac{12\rho_{r}+3\rho_{m}}{4\rho_{r}+3\rho_{m}}\mathcal{H}^{2}%
\right] \Delta_{\pi} &=& k^{2}\Delta_{p}\ \ \ \   \notag
\end{eqnarray}
where $\rho _{m}$ and $\rho _{r}$ are respectively the energy densities of
non-relativistic (baryons plus EDM) and relativistic (photons plus
neutrinos) matter. Here note that the time evolution of $\Delta _{\pi }$ is
driven by $\Delta _{p}$, which itself evolves in accord with Eq.~(\ref%
{eq:EvolveDensityPerturb}), which is driven by $\Delta _{\pi }$ \emph{plus}
the \emph{parametrized} $\Delta _{\mathrm{EDM}}(a)$ (with its time
derivatives). So, in the numerical calculation we have four more variables
to propagate: $\Delta _{p},~v_{\mathrm{EDM}},~\Delta _{\pi }$ and $\Delta
_{\pi }^{\prime }$.

\begin{figure}[tbp]
\centering \includegraphics[scale=0.9] {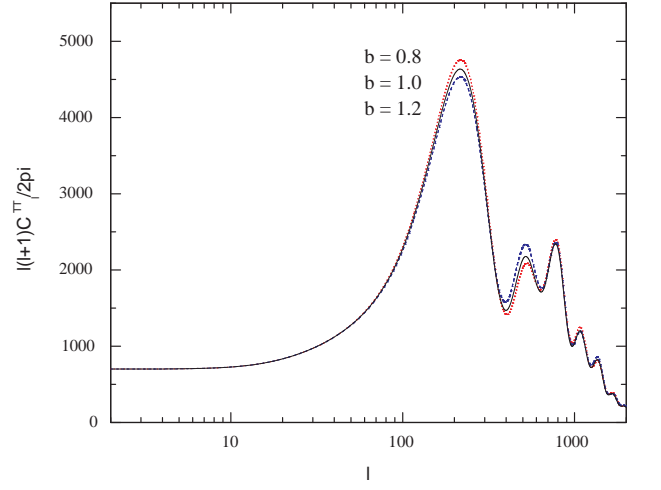}
\caption{(color online) The CMB spectrum for the general case [c.f.~Eq.~(%
\protect\ref{eq:case3Delta_pi})] with $\Delta_{\mathrm{EDM}}(a)$
parameterized as in Eq.~(\protect\ref{eq:EDMparametrization}).
Here we have chosen $a_{i} = 0.0002$ and $a_{f} = 0.002$. The blue
dashed, black solid and red dotted curves represent the cases for
$b=1.2, 1.0$ and $0.8$ respectively, as also shown beside the
curves. The large-angle ($\ell < 100$) CMB powers are
indistinguishable for the curves as in
Fig.~\protect\ref{fig:Figure6} due to the identical evolutions of
$\protect\phi$ and identical ISW effects.} \label{fig:Figure8}
\end{figure}

\begin{figure}[tbp]
\centering \includegraphics[scale=0.9] {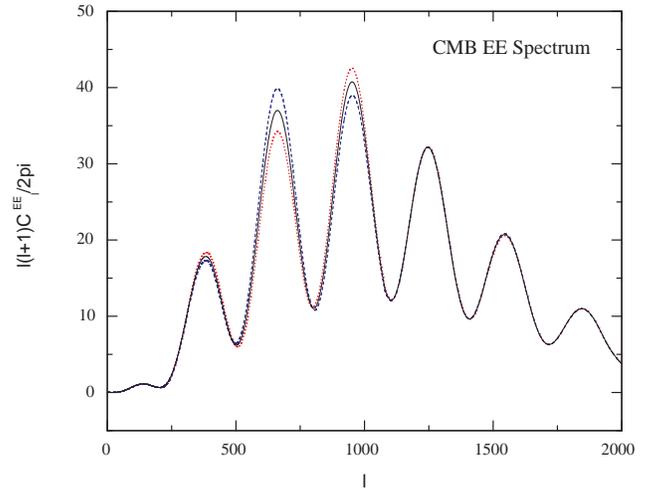}
\caption{(color online) The CMB polarization spectrum for the
model with the same parameters as in
Fig.~\protect\ref{fig:Figure8}. The red dotted, black solid and
blue dashed curves are the cases $b = 0.8, 1.0, 1.2$
respectively.} \label{fig:Figure9}
\end{figure}

Note that if $\Delta_{p}$ vanishes identically and $\Delta_{\pi}=0$
initially then $\Delta_{\pi}$ will remain zero all the time; also if $%
\Delta_{\pi}$ vanishes identically then so does $\Delta_{p}$. These
restrictions indicate that the two cases we have considered in above
subsections \emph{cannot} give rise to the same evolution of $\phi$ \emph{and%
} different evolution of $\Delta_{\mathrm{EDM}}$ to that predicted by
standard CDM at the same time. The general case with $\Delta_{p},~\Delta_{%
\pi} \neq 0$, however, is able to do this, as is shown in Fig.~\ref%
{fig:Figure6}. There, we plot the CMB power spectra of this general case
with different values of $b =1.1,1.0,0.9$ respectively, and they are totally
indistinguishable because the evolutions of $\phi$ are identical, leading to
identical (zero) ISW effects. In this figure we have chosen $a_{i}=0.005$
and $a_{f}=1$ so that we only require the EDM to deviate from CDM after last
scattering. It must be emphasized that, although the three curves are
indistinguishable from each other, they \emph{do} correspond to different
EDM properties. To show this point clearly, we display in Fig.~\ref%
{fig:Figure7} the quantities $c_{s}^{2} \equiv \Delta_{p}/\Delta_{\mathrm{EDM%
}}$ and $c_{v}^{2}\equiv \Delta_{\pi}/\Delta_{\mathrm{EDM}}$, which
characterize the importance of $\Delta_{p}$ and of $\Delta_{\pi}$
respectively, for the models. It is obvious that in the $b \neq 1$ models
these quantities could be very different from the standard CDM value (0),
and this can be understood intuitively as follows: the pressure perturbation
term $\Delta_{p}$ acts \emph{against} gravitational collapse, when $b >1$,
meaning that $\Delta_{\mathrm{EDM}}$ grows faster than $\Delta_{\mathrm{CDM}%
} $ early on and more slowly later; $\Delta_{p}$ needs to be negative early
on and positive later (and vice versa). The anisotropic stress term $\Delta_{%
\mathrm{\pi}}$ dissipates fluctuations in $\Delta _{\mathrm{EDM}}$
[c.f.~Eq.~(\ref{eq:EvolveDensityPerturb})] and behaves similarly in the
figure.

We have thus seen that the observations of the CMB and matter power spectrum
\emph{cannot} rule out this general case (as long as $a_{i}$ is after last
scattering). Cross correlating ISW with the galaxy distribution might help
in this regard, but it is still limited for two reasons: first, it is
contaminated by the effects of dark energy, and second, the
cross-correlation data only exists for late times and cannot effectively
constrain models like ours where the deviation from CDM occurs much earlier.

If we choose the $a_{i}\lesssim 10^{-3}$, which means that the deviation of
EDM from CDM starts before last scattering, then although Eq.~(\ref%
{eq:case3Delta_pi}) guarantees that the $\phi$ evolution is not changed for
arbitrary $b$, the CMB power spectrum will generally be different from that
of standard CDM because the quantities $\Phi$ and $\Psi$, which are directly
relevant for the primary CMB anisotropy, are modified since $\Phi =-\phi -%
\frac{\kappa \Pi a^{2}}{2k^{2}},~\Psi =\phi -\frac{\kappa \Pi a^{2}}{2k^{2}}$%
. In Fig.~\ref{fig:Figure8} we present such an example, for which we have
used the parametrization Eq.~(\ref{eq:EDMparametrization}) of $\Delta_{%
\mathrm{EDM}}(a)$ with $a_{i}=0.0002$ and $a_{f}=0.002$. This shows that in
general the small-angle (high-$\ell $) CMB power will be different even
though the large-angle power is fixed to be the same as in standard CDM. If
the evolution of $\Delta_{\mathrm{EDM}}$ prior to last scattering is
significantly different from that of CDM, then the deviation of CMB spectrum
might be very large, and this is why the CMB could efficiently constrain
such alternative gravitational dark matter theories as TeVeS and the $f(K)$
models.

Another possible constraint comes from the CMB polarization. The
linear polarization of CMB photons carries information about the
gravitational potential because it can only be generated from the
quadrupole moment of the CMB temperature perturbation, which is
related to the anisotropic photon stress \cite{Hu1997a, Hu1997b},
through Thomson scattering. Furthermore, at earlier times the
rapid scattering of photons by electrons ruins the photon
anisotropic stress (c.f.~\S ~\ref{sect:CMB}), consequently the
polarization only appear close to last scattering when significant
photon quadrupole anisotropy can be produced. The localization of
the generation in time, and the dependence on the quadrupole
moment, only mean that the polarization directly reflects the
gravitational potentials at the time of last scattering in a
special way and so provides invaluable information about the EDM.
This is in contrast to the CMB temperature spectrum which depends
on the gravitational potentials through the whole cosmic history
up to now, and also on the photon monopole and dipole moments,
which complicate the extraction of information. In
Fig.~\ref{fig:Figure9} we plot the CMB EE polarization spectrum
for the same model as in Fig.~\ref{fig:Figure8}. It can be seen
that different parameters give quite distinct peak features. The
CMB polarization was first detected in 2002 \cite{DASI}, with the
precision gradually improved since then \cite{Bischoff2008}.
Although the precision at present is still insufficient to place
stringent constraints on the EDM model, we expect that future
observations will change this situation.

We could also slightly change Eq.~(\ref{eq:case3Delta_pi}) to make the
evolution of $\Phi$ \emph{or} $\Psi$ the same as that for standard CDM,
however clearly this is not possible for \emph{both} $\Phi$ \emph{and} $\Psi$
because in this case $\Delta_{\pi}\neq 0$. Because the primary CMB
anisotropy is determined by both of these two variables, it will almost
definitely differ between the EDM and CDM models. Furthermore, in the most
general cases the evolution of $\phi$ will be changed as well, which gives
rise to different low-$\ell$ CMB spectra. Therefore, we expect that the CMB
data will place stringent constraints on the general EDM models, and can be
used to distinguish alternative gravitational theories of dark matter.
Conversely, any attempt of modification of general relativity which claims
to have reproduced the observed large scale structure must be confronted
with the CMB fluctuation spectrum and polarization to test its viability.

\section{Summary and Discussion}

\label{sect:Conclusion}

Motivated by the recent developments in producing the large-scale structure
formation with alternative theories of gravity, we have considered the
prospect of using the CMB to constrain such theories in new ways. We take
the confrontation with the matter power spectrum data as a first test of the
perturbed cosmological model in our alternative gravity theory and assume
that this test has been passed. This is because the currently successful
theories like TeVeS and $f(K)$ really are \emph{only} able to reproduce the
observed LSS rather than all the perturbation observables, and such a
simplifying assumption efficiently restricts our model space so that the
degeneracy problem is somewhat alleviated. As discussed in \S ~\ref%
{sect:Model}, reproducing the observed LSS is a rather weak
requirement on the theory because the LSS only reflects the
density perturbations at late time. This means that there is still
considerable freedom to choose the evolution history of the
dark-matter density perturbation $\Delta_{\mathrm{EDM}}$, which
will generally lead to different predictions about other
cosmological observables such as the CMB spectra and polarization.

We discussed how this freedom in the evolution history can be utilized by
considering an example of the parametrization [c.f.~Eq.~(\ref%
{eq:EDMparametrization})] of $\Delta _{\mathrm{EDM}}(a)$ which only reduces
to the familiar $\Delta _{\mathrm{CDM}}(a)$ result with the parameter choice
$b=1.0$. As was shown in \S ~\ref{sect:Model}, the evolution of $\Delta _{%
\mathrm{EDM}}$ is completely governed by the EDM stress history, which we
quantify using the two variables $\Delta _{p}$ and $\Delta _{\pi }$. These
same two variables also control the evolution of gravitational potential,
which is important for the CMB power spectrum. This implies that the CMB is
an ideal test bed for EDM models. In reality, we have one more freedom
because once the $\Delta _{\mathrm{EDM}}(a)$ is specified. Eq.~(\ref%
{eq:EvolveDensityPerturb}) becomes a single equation for the two variables $%
\Delta _{p}$ and $\Delta _{\pi }$, so, in \S ~\ref{sect:Numerics}, we
considered three separate cases: (I) $\Delta _{\pi }=0,~\Delta _{p}\neq 0$,
(II) $\Delta _{\pi }\neq 0,~\Delta _{p}=0$ and (III) $\Delta _{\pi }\neq
0,~\Delta _{p}\neq 0$. For simplicity, we have also made the following
assumptions: (1) the evolution of $\Delta _{\mathrm{EDM}}$ is scale
independent, as is implied by the observed LSS, (2) the cosmological
constant (or other form of explicit dark energy violating the strong energy
condition) is neglected to simplify the numerics, (3) the EDM equation of
state parameter, $w_{\mathrm{EDM}},$ is assumed to be small enough so that
we can assume the standard CDM background evolution holds; and, (4) only
adiabatic initial conditions, a scale-independent $n_{s}=1$ primordial power
spectrum, and effectively massless neutrinos are considered (we will return
to these assumptions below).

For case I, the standard CDM relation $\Psi =-\Phi =\phi $ still holds but
the evolution of these potentials could be changed. If the deviation of EDM
from CDM only occurs after the last scattering, then this change in $\phi $
mainly modifies the ISW effect and, hence, the low-$\ell $ CMB power. If the
EDM starts to deviate from CDM before last scattering, however, then the
early changes in $\Psi $ and $\Phi $ would modify the CMB acoustic peak
features in a complicated manner, as we described qualitatively in \S ~\ref%
{sect:CMB} and \S ~\ref{subsect:case1}. The evolution of $\Delta _{p}$ can
be understood qualitatively: since the effect of the pressure perturbation
is to counteract gravitational collapse, in order that $\Delta _{\mathrm{EDM}%
}$ grows faster than $\Delta _{\mathrm{CDM}}$ we need a negative $\Delta
_{p},$ and vice versa (c.f.~Fig.~\ref{fig:Figure7}).

The case II evolution is problematic as long as we stick to the above
simplifying assumption (1), \emph{i.e.}, requiring a scale-independent
growth of $\Delta _{\mathrm{EDM}}$, because in this case Eq.~(\ref%
{eq:EvolveDensityPerturb}) implies a scale \emph{dependent} $\Delta _{\pi }$
($k^{2}\Delta _{\pi }$ is independent of $k$), which diverges on large
scales (small $k$). As we have seen in \S ~\ref{subsect:case2}, this leads
to a strong late-ISW effect, blowing up the low-$\ell $ CMB spectrum even if
the EDM only differs slightly from standard CDM . Furthermore, a natural EDM
model with $\Delta _{p}=0$ and $\Delta _{\pi }\neq 0$ does not exist in the
literature, and this model is different from of case I, which arises
naturally in $f(K)$ models with $c_{13}=0$.

Of course, the most general situation is our case III, where both $\Delta
_{p}$ and $\Delta _{\pi }$ are nonzero. Needless to say, this generality
also makes its exploration more difficult. One could, however, use the extra
degree of freedom here in model constructions. If we fix \emph{both} the
evolution of $\Delta _{\mathrm{EDM}}(a)$ \emph{and} that of the potential $%
\phi $, then $\Delta _{\pi }$ and $\Delta _{p}$ can be solved uniquely at
the same time. Such a construction has the advantage that we can fix the
evolution of $\phi $ to be identical to that in standard CDM so that the ISW
effect is not changed at all. If the deviation of EDM starts after last
scattering, then such a model is degenerate with standard CDM even after CMB
and LSS data are taken into account, and furthermore the CMB-galaxy cross
correlation data is inefficient in constraining it. Allowing EDM to deviate
before last scattering, however, will almost surely predict different CMB
peak features, because in this case $\Psi \neq -\Phi \neq \phi $ and the
acoustic oscillations of photon-baryon fluid are changed with respect to
standard CDM (c.f.~\S ~\ref{sect:CMB}). Dropping the requirement on the
evolution of $\phi $ makes the situation even worse, because in this case
the ISW effect and low-$\ell $ structure of the CMB spectrum are also
changed. We stress that the CMB polarization also provides invaluable
information for constraining EDM models, because it depends only on the
quadrupole moment of the photon temperature perturbation $\Theta $ (the
photon anisotropic stress), and thus only on the physics close to the time
of last scattering.

Throughout this paper we have chosen to parametrize $\Delta _{\mathrm{EDM}%
}(a)$ as in Eq.~(\ref{eq:EDMparametrization}). This is fairly simple and
sufficient for our purposes as we only aim to show the importance of CMB
data in constraining alternative gravitational dark matter theories, but not
to constrain any specified model or given parametrization exactly. There are
good reasons why more detailed parametrizations are needed for more precise
future studies. First, although $\Delta _{\mathrm{CDM}}\propto a$ is a good
approximation in the matter-dominated epoch, it breaks down when the
contribution from radiation is still significant (including the time prior
to and around last scattering) and when a cosmological constant is included.
Second, because the physics at the time of matter-radiation equality is
relevant for EDM models which deviate from CDM earlier, we sometimes need to
choose $a_{i}$ to be smaller than the matter-radiation equality $a_{\mathrm{%
eq}}$, which means that the evolution of $\Delta _{\mathrm{EDM}}(a)$ or $%
\Delta _{\mathrm{CDM}}(a)$ cannot be completely scale independent (remember
that $a_{\mathrm{eq}}$ is relevant for the bending of the matter power
spectrum \cite{LMB 1998}). Thus, future precise calculations, particularly
those relevant for weak lensing and the galaxy-ISW cross correlation, should
use parametrizations which reduce exactly to $\Delta _{\mathrm{CDM}}(a)$ in
the $\Lambda $CDM model in appropriate limits.

Notice that our logic here is different from other related considerations of
generalized dark matter (GDM) \cite{Hu1999, Hu1998}. In Ref.~\cite{Hu1998},
for example, the author parametrizes the stress sector of the GDM with three
quantities, $w_{g},~c_{eff}^{2}$ and $c_{vis}^{2}$, which respectively
characterize the EOS, pressure perturbation, and anisotropic stress of the
GDM. The GDM effects on the CMB \emph{and} matter power spectra are then
studied by assuming some specific values of these quantities. Our approach
tackles the problem from a different direction and is therefore
complementary to those earlier works.

One interesting issue is the inclusion of hot dark matter, which is needed
by TeVeS itself. This can certainly be achieved by setting special values
for $w_{g},~c_{eff}^{2}$ and $c_{vis}^{2}$ as in \cite{Hu1998}. In our
approach we simply treat the EDM as a "black box" which is completely
controlled by separate conservation equations. By parametrizing $\Delta _{%
\mathrm{EDM}}(a)$ we are able to extract information on the pressure
perturbation $\Delta _{p}$ and anisotropic stress $\Delta _{\pi }$ from
these conservation equations without knowing exactly what is in the box --
it can be mixtures of modified gravitational effects and neutrinos, or
something else entirely.

The other issue is related to the initial conditions. In this work we mainly
focus on the adiabatic initial conditions as in \cite{Hu1998}. However, it
is well known that the adiabatic initial condition is far from the only
possibility, and there can be four regular isocurvature modes \cite%
{Bucher2000}. These isocurvature modes are predicted by many theoretical
models and may even correlate with the adiabatic one. In contrast to the
adiabatic mode, the isocurvature mode excites sinusoidal oscillations of the
photon-baryon fluid and so predicts the first CMB acoustic peak to be around
$\ell \sim 330$ rather than $220$; consequently a pure or dominating
isocurvature initial condition is incompatible with basic CMB observations.
Nonetheless, a subdominant contribution from isocurvature initial conditions
actually degenerates with other cosmological parameters and cannot be ruled
out by the current data \cite{Trotta2001, Amendola2002, Beltran2004,
Keskitalo2007}. In our EDM model, if the EDM evolves differently from CDM
early in the radiation era, then $\Delta _{\mathrm{EDM}}$ does not evolve
adiabatically and the entropy perturbation $S_{\mathrm{EDM}}=\Delta _{%
\mathrm{EDM}}-\frac{3}{4}\Delta _{\gamma }$ is nonzero. In this case, one
might naturally expect that there should be isocurvature modes in the
initial conditions. A detailed calculation of the consequences of
introducing such modes, however, generally requires thorough searches of the
new parameter space (which is enlarged compared to the standard case because
now the amplitude, tilt of the isocurvature modes and their correlation with
the adiabatic mode must be taken as free parameters) like in \cite%
{Trotta2001, Amendola2002, Beltran2004, Keskitalo2007} and is beyond the
scope of this work. Moreover, allowing different initial conditions
(especially tilts of the isocurvature mode which are significantly different
from 1) can change the shape of matter power spectrum, which means that the
parametrization of $\Delta _{\mathrm{EDM}}(a)$ [c.f.~Eq.~(\ref%
{eq:EDMparametrization})] should be scale dependent for the same reasons as
discussed above. In this work we adopt a more conservative approach by
assuming that the EDM evolves like standard CDM at earlier times and so
adiabaticity is a natural choice.

In conclusion, we propose ways to use the CMB in order to constrain those
alternative gravity theories for dark matter which claim to be compatible
with \ the observed LSS. We find that the CMB temperature and polarization
spectra are good discriminators between these theories in general,
especially when they deviate from the CDM paradigm before last scattering.
If the deviation starts after last scattering, however, there can exist EDM
theories which are degenerate with respect to the standard $\Lambda $CDM
model and cannot be distinguished by CMB and matter power spectra. This
point is particularly interesting from the viewpoint of model constructions.
Our results also indicate that the stress properties of dark matter, which
determine the evolutions of both density perturbations and gravitational
potential, can be studied and significantly constrained with existing and
future data by using just the general conservation equations and without
specialising to any specific theoretical model.


\begin{acknowledgments}
We thank David Spergel for helpful discussions. The numerical calculation of
this work uses a modified version of the CAMB code \cite{CAMB}. BL thanks
University of St.~Andrews for its hospitality when part of this work was
carried out and acknowledges supports from an Overseas Research Studentship,
the Cambridge Overseas Trust, the DAMTP and Queens' College at Cambridge.
DFM acknowledges the Humboldt Foundation.
\end{acknowledgments}

\end{document}